\documentclass[12pt]{iopart}
\usepackage{graphicx}
\usepackage{subfigure}
\usepackage{setspace}
\usepackage{cite}
\usepackage{epstopdf}
\begin{document}

\title[The twin paradox in curved spacetime]{A computational approach to the twin paradox in curved spacetime}

\author{Kenneth K H Fung$^{1,2,3}$, Hamish A Clark$^{1,4}$, Geraint F Lewis$^{1,5}$ and Xiaofeng Wu$^{2,6}$}
\address{$^1$Sydney Institute for Astronomy, School of Physics, A28, The University of Sydney, NSW 2006, Australia}
\address{$^2$School of Aerospace, Mechanical and Mechatronic Engineering, J07, The University of Sydney, NSW 2006, Australia}
\ead{$^3$\mailto{kfun2342@uni.sydney.edu.au}, $^4$\mailto{hamish.clark@sydney.edu.au}, $^5$\mailto{geraint.lewis@sydney.edu.au}, $^6$\mailto{xiaofeng.wu@sydney.edu.au}}
\vspace{10pt}
\begin{indented}
\item[]June 2016
\end{indented}

%---------------------------------------------------------------------------------------------------------------

\begin{abstract}
Despite being a major component in the teaching of special relativity, the twin `paradox' is generally not examined in courses on general relativity. Due to the complexity of analytical solutions to the problem, the paradox is often neglected entirely, and students are left with an incomplete understanding of the relativistic behaviour of time. This article outlines a project, undertaken by undergraduate physics students at the University of Sydney, in which a novel computational method was derived in order to predict the time experienced by a twin following a number of paths between two given spacetime coordinates. By utilising this method, it is possible to make clear to students that following a geodesic in curved spacetime does not always result in the greatest experienced proper time.
\end{abstract}

\pacs{04.20.-q, 03.30.+p, 95.30.Sf}
\vspace{2pc}
\noindent{\it Keywords\/}: Relativity, Twin Paradox, Black Holes, Wormholes \\
\submitto{\EJP}
%\maketitle

%---------------------------------------------------------------------------------------------------------------

\section{Introduction}
The twin `paradox' has experienced countless re-imaginings and interpretations throughout the course of relativistic physics, and has a long and controversial history \cite{frisch, builder, scott}. In its most basic form, it is often the first description of non-absolute time that a student will encounter. It acts as a fantastic opportunity to test their understanding of special relativity, with each variation upon the paradox providing a new viewpoint on the `relative' nature of time.

In its original form, the paradox may be stated as follows: There are two identical twins -- one of which climbs aboard a rocket and travels away at relativistic speeds, while the other remains on Earth. After a while, the travelling twin turns around and returns home. Each twin has seen the other as moving away and returning, and so a naive application of special relativistic time dilation will predict that each twin would expect themself to be older upon reuniting. The resolution to this may be seen by recognising that the experience of each twin is not entirely identical -- the travelling twin has to turn around and return to Earth, while the other remains in an inertial frame throughout. Although further extensions to the paradox have been made in an attempt to remove acceleration, the same conclusion is always found: in flat spacetime, the stationary twin is always eldest upon reunion, whilst the travelling twin will always be younger. In other words, the twin who follows a geodesic path (i.e. remains in an inertial frame) will experience a greater proper time between two events in flat spacetime.

An infrequently examined form of the paradox may be seen upon extension to the framework of general relativity -- by ensuring that the path of each twin passes through curved spacetime. This extension is able to provide apparently contradictory results to the original, as multiple geodesic paths between events may exist. For example, one may imagine that in the vicinity of a point mass, a twin wishing to return to their original position may do so by either following an orbital path around the mass, or a purely radial one (travelling outward and freefalling back down). Conversely, a twin who is stationary with respect to the mass is no longer in an inertial frame, and must necessarily feel a force due to acceleration. Identifying the eldest twin in this case is no longer trivial -- this question may no longer be answered by simply determining which twin undergoes acceleration. While it is known that the longest path between two points must be a geodesic, it is not immediately obvious which geodesic this must be. Due to the increased complexity of the general relativistic paradox, prior approaches have only investigated the simplest cases in a given spacetime individually, and have done so analytically \cite{swift, abram1, abram2, gron, soko, markley}. 

In what follows, we describe a novel computational method to trace multiple paths between two spacetime events, allowing for the general relativistic twin paradox to be made easily accessible. As an example of the strength and simplicity of this method, we investigate accelerated and geodesic paths in both the Morris-Thorne and Schwarzschild metrics. This method was devised by undergraduate students at the University of Sydney, and may be applied to any chosen metric. The difficulty of its application is set by the complexity of the metric. As such, this method could be utilised equally in the teaching of courses on general relativity.

%---------------------------------------------------------------------------------------------------------------

\section{Method}
In general relativity, a twin in free-fall will follow a path described by a geodesic. In the case that a twin should activate their rocket at any point on their path, their deviation from the geodesic may be described by a 4-acceleration, defined in Einstein summation convention as

\begin{equation}
a^\alpha = u^\mu \nabla_\mu u^\alpha = \frac{d^2x^\alpha}{d\tau^2} + \Gamma_{\beta\gamma}^{\alpha}u^\beta u^\gamma,
\label{geo}
\end{equation}
where $\nabla_\mu$ is the covariant derivative, $\tau$ is the time indicated by a standard clock carried by the twin moving along the path $x^\alpha(\tau)$ with 4-velocity $u^\alpha$, and the Christoffel symbols $\Gamma_{\beta\gamma}^{\alpha}$ are unique to the spacetime through which the twin is passing. We also require that the twins follow a `timelike' path, providing the requirement

\begin{equation}
\mathbf{u\cdot u} = g_{\alpha\beta}u^{\alpha}u^{\beta}=-1,
\label{u.u}
\end{equation}
where $g_{\alpha\beta}$ is the metric tensor of the specific spacetime, and the signature convention $(-+++)$ has been adopted. We have also used geometrised units, where $c=G=1$. \\It follows from (\ref{geo}) that the 4-acceleration of a twin is perpendicular to their 4-velocity:
\begin{equation}
\mathbf{u\cdot a}=g_{\alpha\beta}a^\alpha u^\beta=0.
\label{a.u}
\end{equation}

In order to examine the twin paradox, we require that each path begin and end at the same pair of spacetime coordinates. As such, this problem must be solved as a boundary value problem (BVP), with bounds on $t$ and $x^i$. The equations of motion must be solved in terms of $t$, such that the proper time may simply be output upon reuniting the paths. Rearranging (\ref{geo}) and defining $\nu^\tau \equiv d\tau/dt$, $\nu^t \equiv 1$, and $\nu^i \equiv dx^i/dt$, we find the following equations of motion

\begin{equation}
\frac{d\nu^\tau}{dt} = \nu^\tau\Gamma_{\beta\gamma}^{t}\nu^\beta\nu^\gamma - a^t\left(\nu^\tau\right)^3,
\label{accelgeo1}
\end{equation}
\begin{equation}
\frac{d\nu^i}{dt} = (\nu^i\Gamma_{\beta\gamma}^t - \Gamma_{\beta\gamma}^i)\nu^\beta\nu^\gamma + \left(\nu^\tau\right)^2 \left(a^i - a^t\nu^i\right).
\label{accelgeo2}
\end{equation}

Similarly, the normalisation conditions (\ref{u.u}) and (\ref{a.u}) now become
\begin{equation}
g_{\alpha\beta}\nu^{\alpha}\nu^{\beta}=-(\nu^\tau)^2,
\label{u.u2}
\end{equation}
\begin{equation}
g_{\alpha\beta}a^\alpha \nu^\beta=0.
\label{a.u2}
\end{equation}

In this framework, $a^i$ may be chosen arbitrarily, and the corresponding $a^t$ derived from rearranging (\ref{a.u}), giving:
\begin{equation}
a^t = -\frac{g_{\alpha i}u^\alpha a^i}{g_{\beta t}u^\beta}.
\label{at}
\end{equation}

To investigate geodesic paths, we simply set $a^i = 0$ in (\ref{at}), giving $a^t = 0$. Equations (\ref{accelgeo1}) and (\ref{accelgeo2}), which are a set of four 2nd order ordinary differential equations (ODEs), may now be split into eight 1st order ODEs and solved as a boundary value problem. Matlab's \texttt{bvp4c} solver is used to solve the ODE's, and requires a total of 8 boundary conditions. To reunite the twins, boundaries must be placed on the initial and final coordinates $\nu^\alpha$, where the initial value of $\nu^\tau$ is set by application of (\ref{u.u2}) as
\begin{equation}
\nu^\tau = \left(-g_{\alpha\beta}\nu^\alpha\nu^\beta\right)^{1/2}.
\end{equation}

We now may straightforwardly input the metric tensor and Christoffel symbols of a particular spacetime, specify the initial and final coordinates, and the \texttt{bvp4c} solver will determine the initial derivatives of each variable. If the region between two given spacetime points is numerically tractable, the proper time experienced by each twin will simply be output at the end of their paths.

A sample of the code has been included which allows the reader to reproduce the following results, and is included in \ref{appA} so as to not detract the reader's attention from the flow of the paper. Comments have also been included within the code to aid the reader.

%---------------------------------------------------------------------------------------------------------------

\section{Application to Simple Metrics}
We now apply this method to two simple metrics -- the Morris-Thorne wormhole, and the Schwarzschild black hole. By investigating both accelerated and geodesic paths, we were able to identify the path between two spacetime events that a twin follows to maximise the proper time they experience.

In each of the following scenarios, the twins were reunited after a coordinate time $T$, defined by
\begin{equation}
T=\sigma P=2\pi\sigma\sqrt{\frac{r_0^3}{M}},
\end{equation}
where $r_0$ is the initial radial coordinate distance from the origin, and $P$ is the coordinate time taken to complete a geodesic circular orbit at a radius of $r_0$ around a Schwarzschild black hole of mass $M$. Here, $\sigma$ is some chosen factor that will be varied to set the coordinate time at which the twins reunite. 

For accelerated paths, only the radial component of the four-acceleration is varied, so that $a^\theta=a^\phi=0$. The radial component $a^r$ of each twin is varied linearly as a function of the coordinate time $t$:
\begin{equation}
a^r=\frac{4a_0}{T}\left| t-\frac{T}{2}\right|-a_0,
\label{areqn}
\end{equation}
where $a_0$ is the maximum magnitude of $a^r$, as demonstrated in Figure \ref{arplot} below.

\begin{figure}[h!]
\centering
\includegraphics[width=0.5\textwidth]{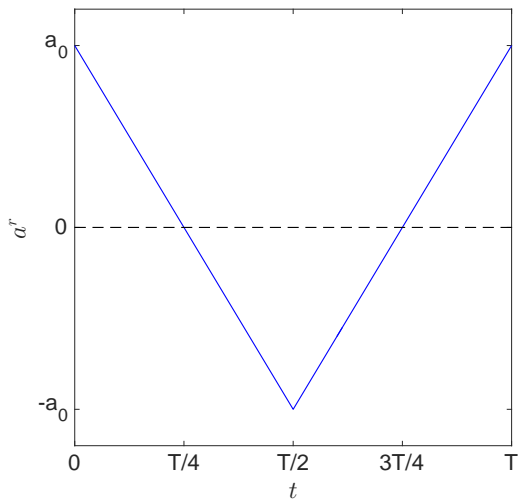}
\caption{Radial component of four-acceleration as a function of the coordinate time.}
\label{arplot}
\end{figure}

%---------------------------------------------------------------------------------------------------------------

\subsection{Morris-Thorne Wormhole}
The Morris-Thorne (MT) metric describes a spherically symmetric, non-vacuum, traversable `wormhole' solution to the Einstein field equations \cite{mt}. This wormhole acts as a link between two seperate regions of spacetime, and requires a distribution of exotic matter with negative energy density. The line element of this metric (Equation B2a in \cite{mt}, Equation 2.16.1 in \cite{catalogue}) is:
\begin{equation}
ds^{2}=-dt^{2}+d\ell^{2}+\left(\ell^{2}+b^2\right)\left(d\theta^{2}+\sin^{2}\theta\,d\phi^{2}\right),
\end{equation}
where $b$ is the wormhole throat parameter, describing the width of the throat, and $\ell$ is the radial proper distance, ranging from $-\infty$ to $+\infty$ on each `side' of the wormhole respectively. Here, we have used a wormhole throat parameter $b = 1$; varying this parameter will not change the results found.

\begin{figure}[h!]
\centering
\subfigure{
\includegraphics[width=0.47\textwidth]{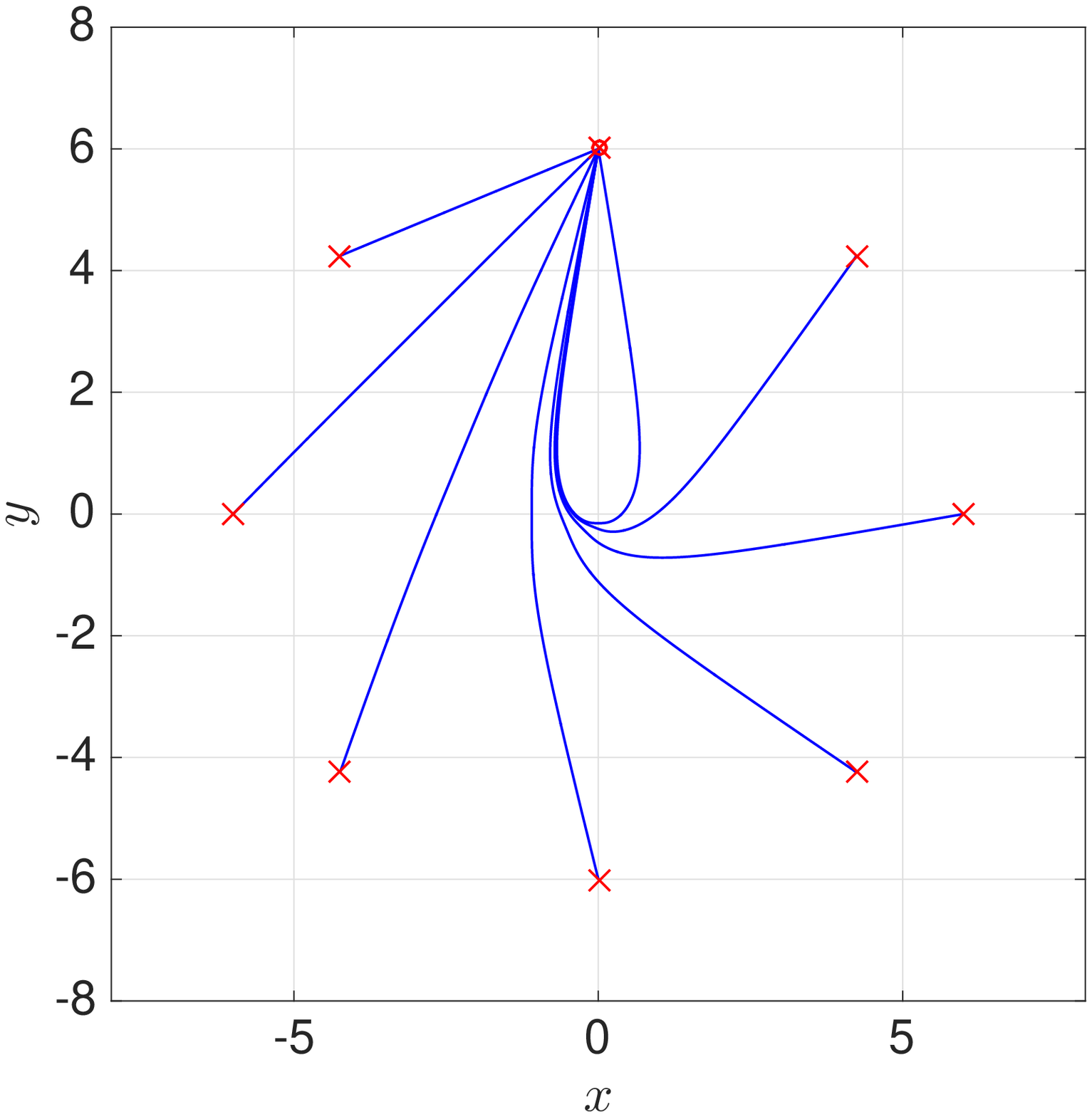}
\label{MTorbit}
}
\subfigure{
\includegraphics[width=0.47\textwidth]{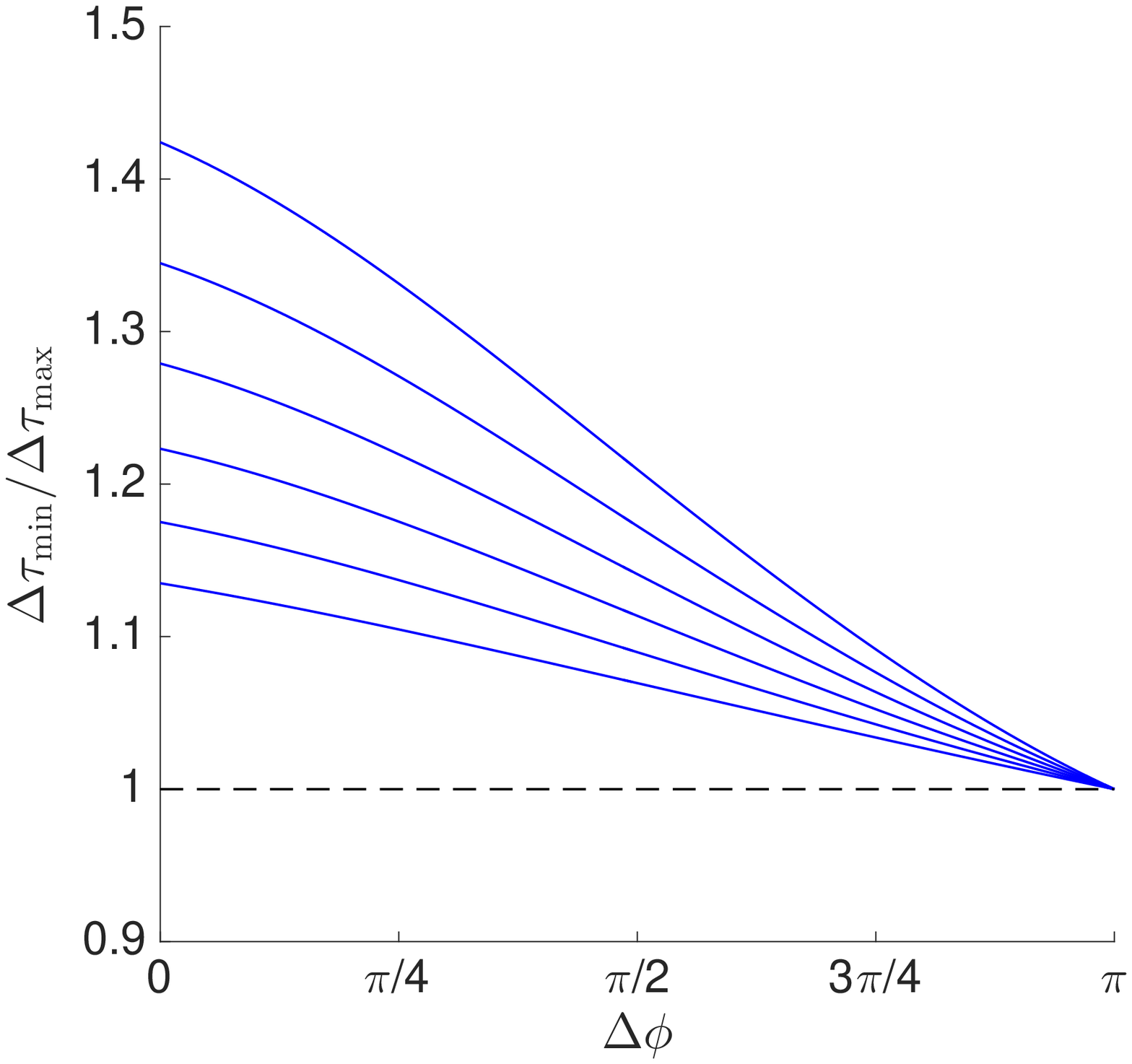}
\label{MTgeo}
}
\caption{(a) A sample of geodesic paths within the MT wormhole metric, where the wormhole throat lies at the origin. The elapsed coordinate time was calculated using $M=1$, $r_0=6$, and $\sigma=0.25$. Time-like geodesics start at $\ell = 6$, $\phi = \pi/2$, $\theta = \pi/2$, and end at $\ell=6$, $\theta=\pi/2$, with 8 angular endpoints $\phi=n\pi/4$ for $n=3,4,\ldots,10$. (b) The proper time ratio of the azimuthally minimal ($\Delta \phi<\pi$) to azimuthally maximal ($\Delta \phi>\pi$) geodesics between various pairs of events within the MT metric. A range of final angular and radial positions are explored for with initial values of $\ell=6$, $\phi=\pi/2$, $\theta=\pi/2$, and ending at $\theta=\pi/2$. The elapsed coordinate time was calculated using $M=1$, $r_0=6$, and $\sigma=0.25$. Changes in angular position are plotted along the $x$-axis, with each line corresponding to final radial distances $\ell_f=1,2,\ldots,6$.}
\end{figure}

Upon inputting the Christoffel symbols of the MT spacetime \cite{catalogue} to the method previously outlined, it was found that for any pair of spacetime points on the same `side' of the wormhole, two geodesic paths exist between them -- as shown in Figure~\ref{MTorbit}. The first of these passes close to the wormhole throat without passing through, bending around to reach the geodesic endpoint. The other follows a more direct route, covering a smaller azimuthal angular distance $\Delta\phi$, and is bent only marginally by the wormhole. Direct comparison of the proper time along these paths, as shown in Figure~\ref{MTgeo}, provides an initial resolution: the twin that covers the smallest azimuthal distance without accelerating will be eldest within the MT spacetime.

\begin{figure}
\centering
\subfigure{
\includegraphics[width=0.47\textwidth]{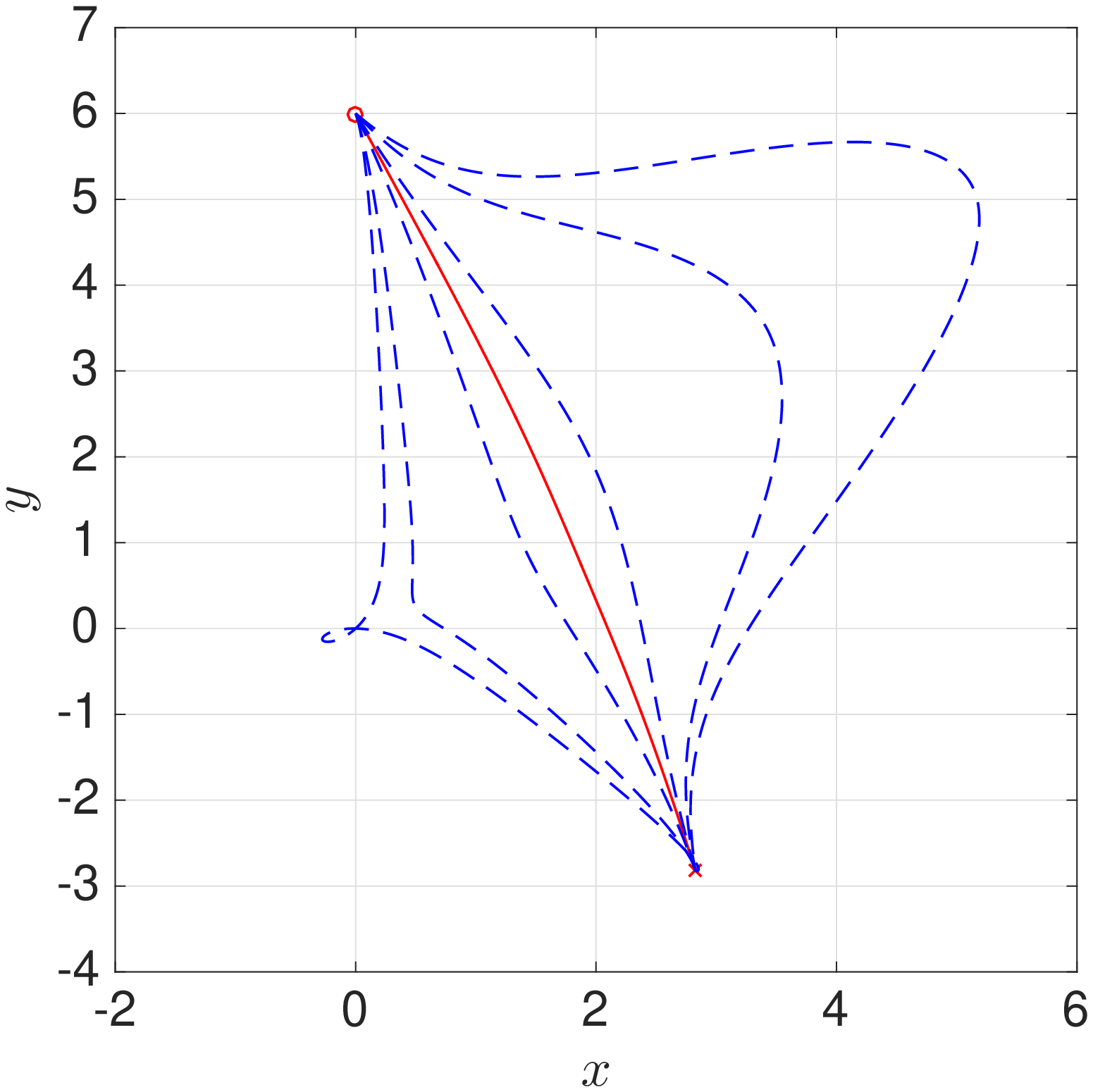}
\label{MTorbitaccel}
}
\subfigure{
\includegraphics[width=0.47\textwidth]{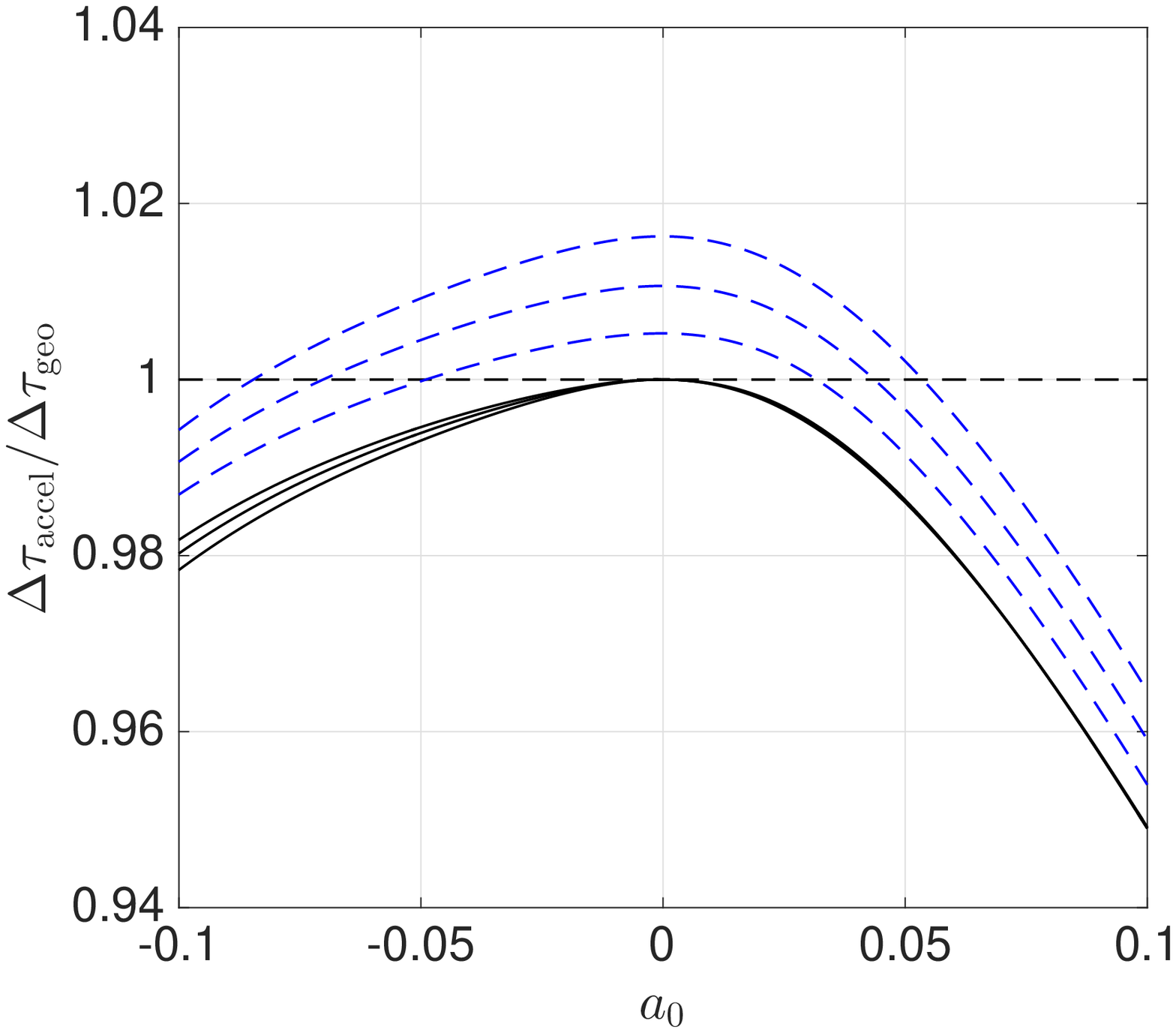}
\label{MTaccel}
}
\caption{(a) A geodesic path (solid) and a sample of accelerated paths (dashed) within the MT wormhole metric, where the wormhole throat lies at the origin. The elapsed coordinate time was calculated using $M=1$, $r_0=6$, and $\sigma=0.4$. Time-like paths start at $\ell=6$, $\phi=\pi/2$, $\theta=\pi/2$, and end at $\ell=4$, $\phi=-\pi/4$, $\theta=\pi/2$. A range of values for $a_0$ were used: $a_0=\pm0.01,\pm0.05,\pm0.1$. (b) Proper time ratios for accelerated paths to geodesic paths within the MT metric. The elapsed coordinate time was calculated using $M=1$, $r_0=6$, and $\sigma=0.4$. Time-like paths begin at $\ell=6$, $\phi=\pi/2$, $\theta=\pi/2$, and end at $\ell=4$, $\theta=\pi/2$. A range of final angular positions are explored: $\phi = 39\pi/30,41\pi/30,43\pi/30$. The solid lines correspond to geodesics that traverse the smaller azimuthal distance, while the dashed lines correspond to geodesics which traverse the non-minimal azimuthal distance.}
\end{figure}

Examination of accelerated paths gives greater insight into the nature of each set of geodesics, as shown in Figure~\ref{MTorbitaccel}. The radial component of four-acceleration of each path is varies according to (\ref{areqn}). By exploring a range of values for the maximum radial component of the four acceleration, $a_0$, we can compare the proper time experienced along each of these accelerated paths to that experienced along geodesics -- the ratio of these proper times is plotted in Figure~\ref{MTaccel}. Interestingly, we find accelerated paths that have proper times greater than the geodesic that covers the larger azimuthal distance. In contrast to the special relativistic paradox, we find that acceleration will not always cause a twin to be younger. Rather, there appears to be a hierarchy of the experienced proper times, in decreasing order as: azimuthally minimal geodesics, `weakly' accelerated paths, azimuthally maximal geodesics, and `strongly' accelerated paths.

%---------------------------------------------------------------------------------------------------------------

\subsection{Schwarzschild Black Hole}
The spacetime geometry in the vicinity a non-rotating and uncharged black hole of mass $M$ may be described by the line element of the Schwarzschild metric in Eddington-Finkelstein coordinates (Equations 2.2.35, 2.2.36 in \cite{catalogue}):

\begin{equation}
ds^2=-\left(1-\frac{2M}{r}\right)dv^{2}+2\,dv\,dr+r^{2}\left(d\theta^{2}+\sin^{2}\theta\, d\phi^{2}\right),
\end{equation}
where
\begin{equation}
v = t + r + 2M\ln\left(\frac{r}{2M}-1\right).
\end{equation}

\begin{figure}[h!]
\centering
\subfigure{
\includegraphics[width=0.47\textwidth]{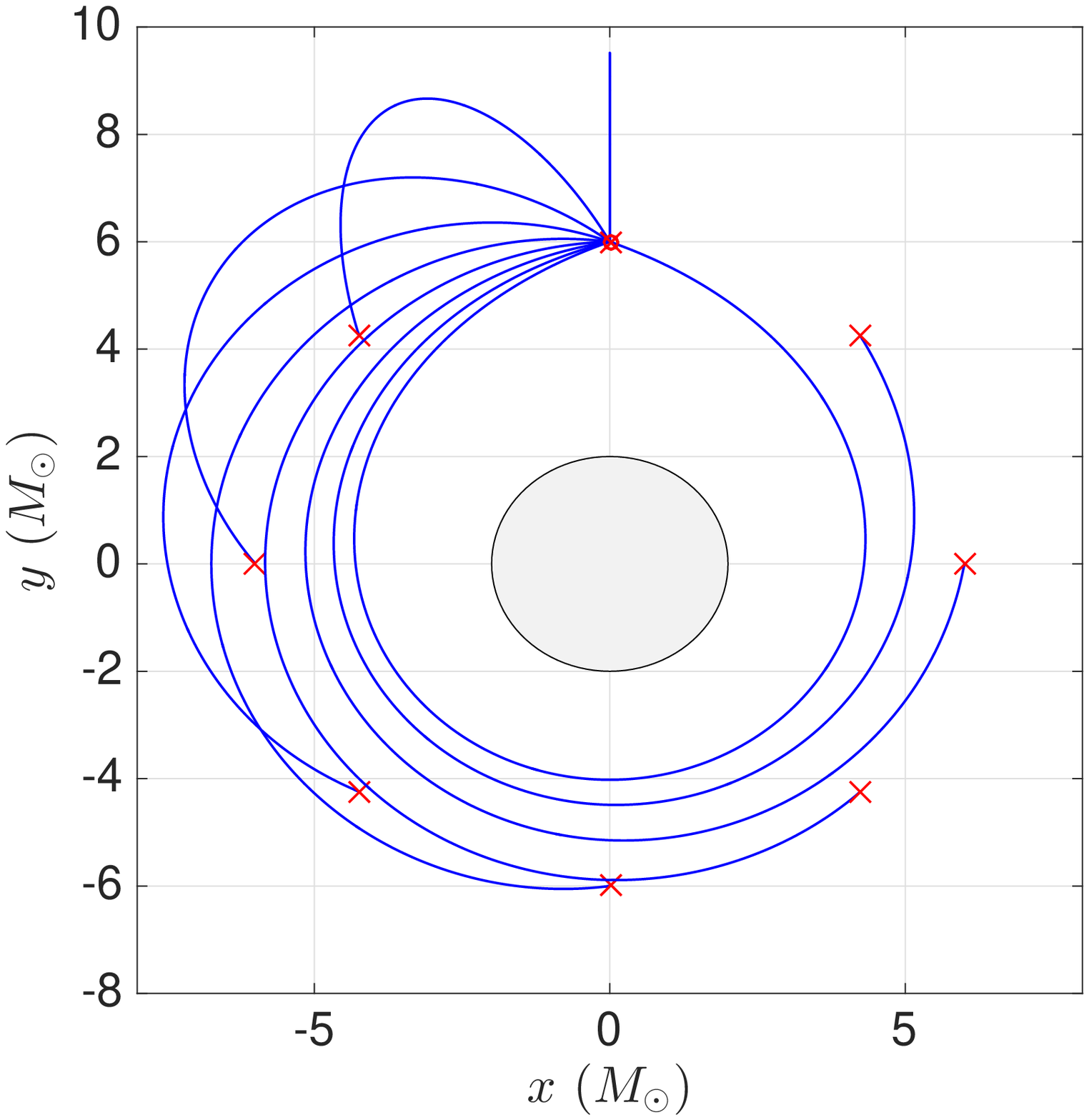}
\label{Schwarzorbits}
}
\subfigure{
\includegraphics[width=0.47\textwidth]{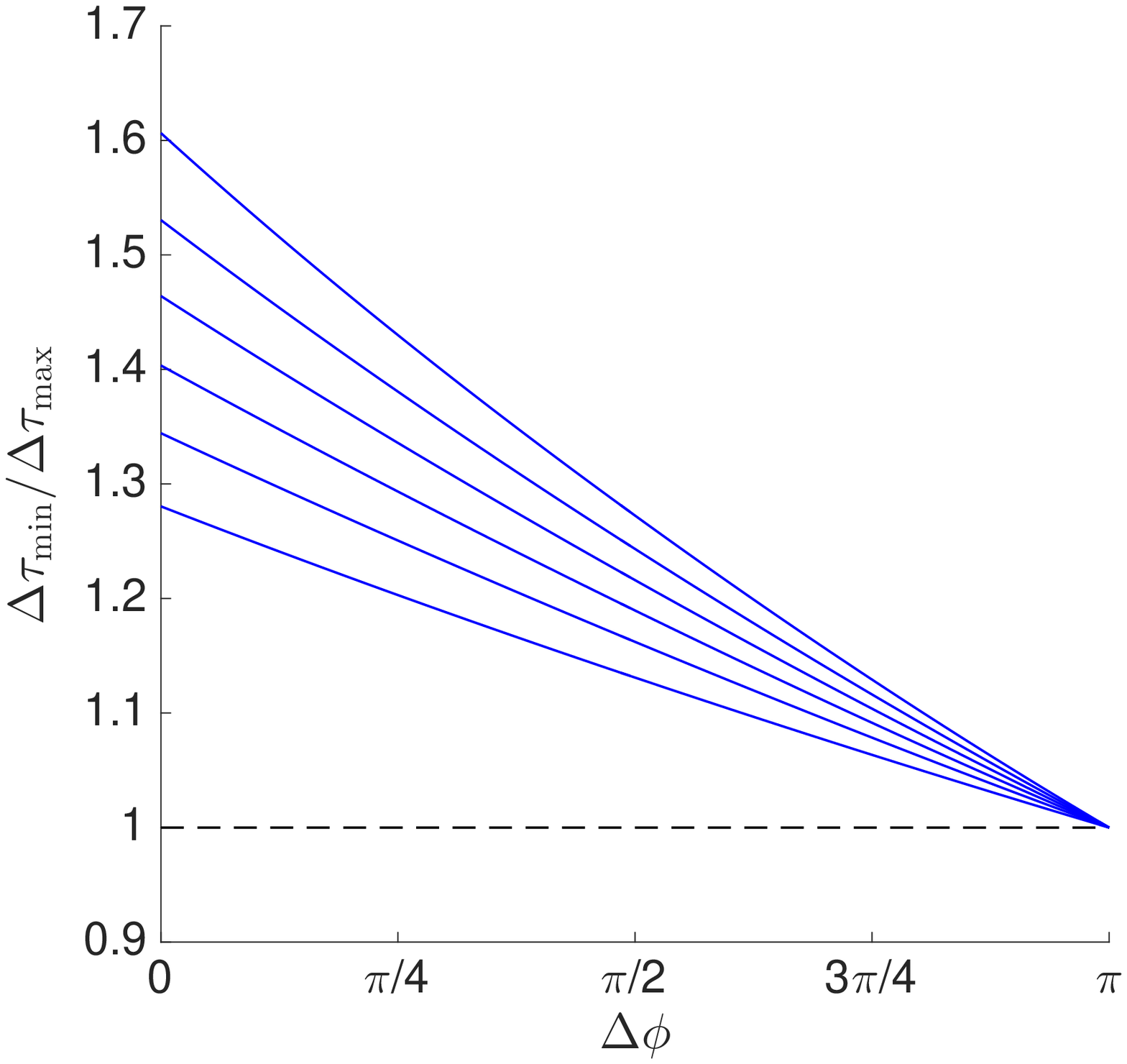}
\label{Schwarzgeo}
}
\caption{(a) A visualisation of a sample of geodesic paths within the Schwarzschild metric, with the Schwarzschild radius $r_s = 2M$ shown in black. The elapsed coordinate time was calculated using $M=1$, $r_0=6$, and $\sigma=0.6$. Time-like geodesics start at $r=6$, $\phi=\pi/2$, $\theta=\pi/2$, and end at $r=6$, $\theta=\pi/2$, with varying angular endpoints on the interval $\pi/2\leq\phi\leq5\pi/2$. Due to the spherical symmetry of the metric, this slice through the black hole will be representative of any set of paths throughout. (b) The proper time ratio of minimal azimuthal geodesics to the maximal azimuthal geodesics between various pairs of events within the Schwarzschild metric. The elapsed coordinate time was calculated using $M=1$, $r_0=6$, and $\sigma=0.6$. A range of final angular and radial positions are explored with initial values $r=6$, $\phi=\pi/2$, $\theta=\pi/2$, and ending at $\theta=\pi/2$. Each line corresponds to final radii $r_f$ both within and outside of the event horizon: $r_f=1,2,\ldots,6$.}
\end{figure}

In this coordinate system, twins may reunite within the event horizon without encountering a coordinate singularity at the Schwarzschild radius, $r_s = 2M$. Here, we have used a black hole mass of $M = 1$; varying this parameter will not change the results found. As with the method followed for the MT wormhole, the Christoffel symbols for this metric were input, and a range of paths compared. Similar to the previous case, two types of geodesics were encountered: those which cover the minimal azimuthal distance, and those covering a non-minimal azimuthal distance -- a number of each of these are shown in Figure~\ref{Schwarzorbits}.

\begin{figure}[h!]
\centering
\subfigure{
\includegraphics[width=0.46\textwidth]{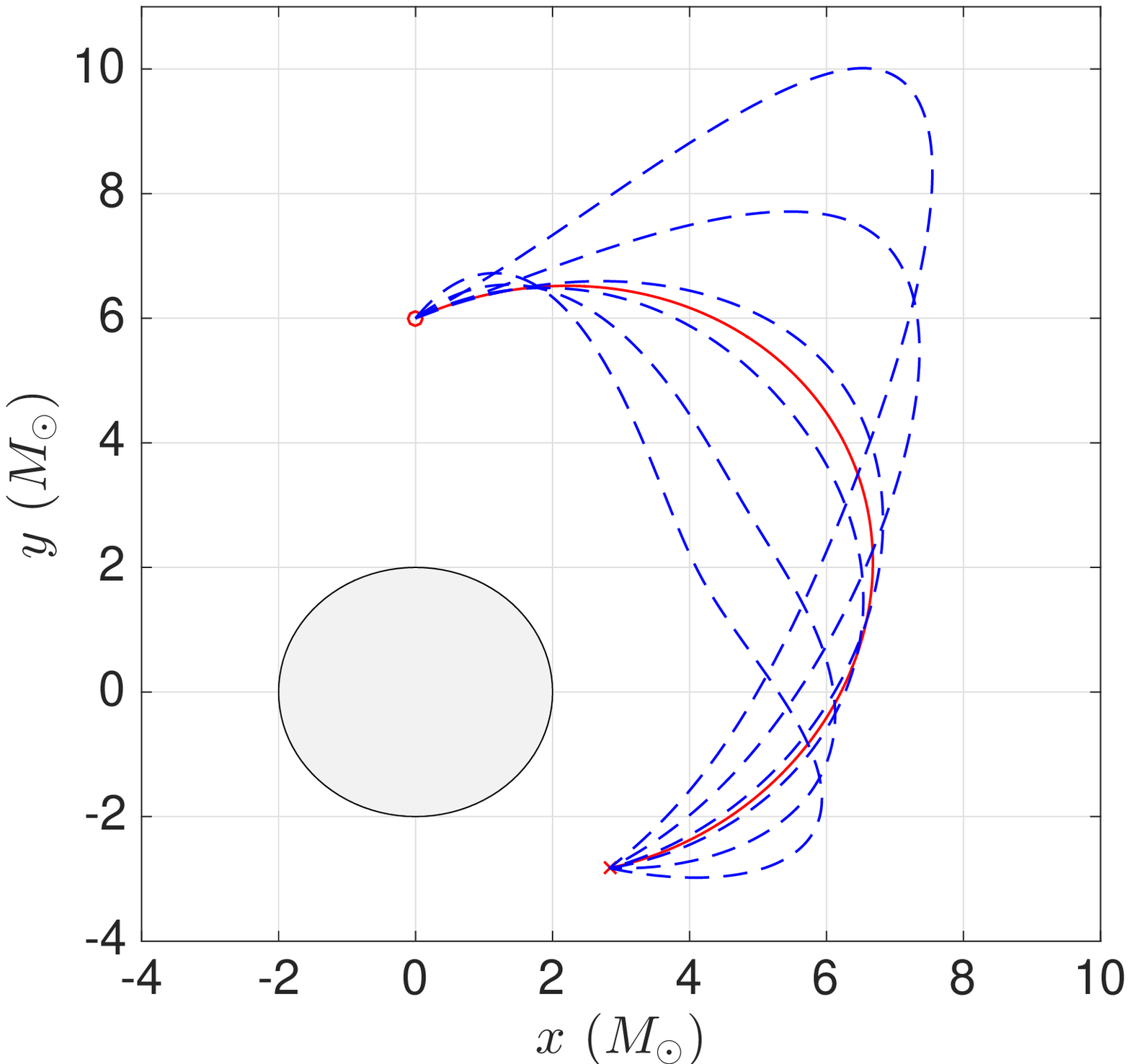}
\label{BHaccelorbit}
}
\centering
\subfigure{
\includegraphics[width=0.46\textwidth]{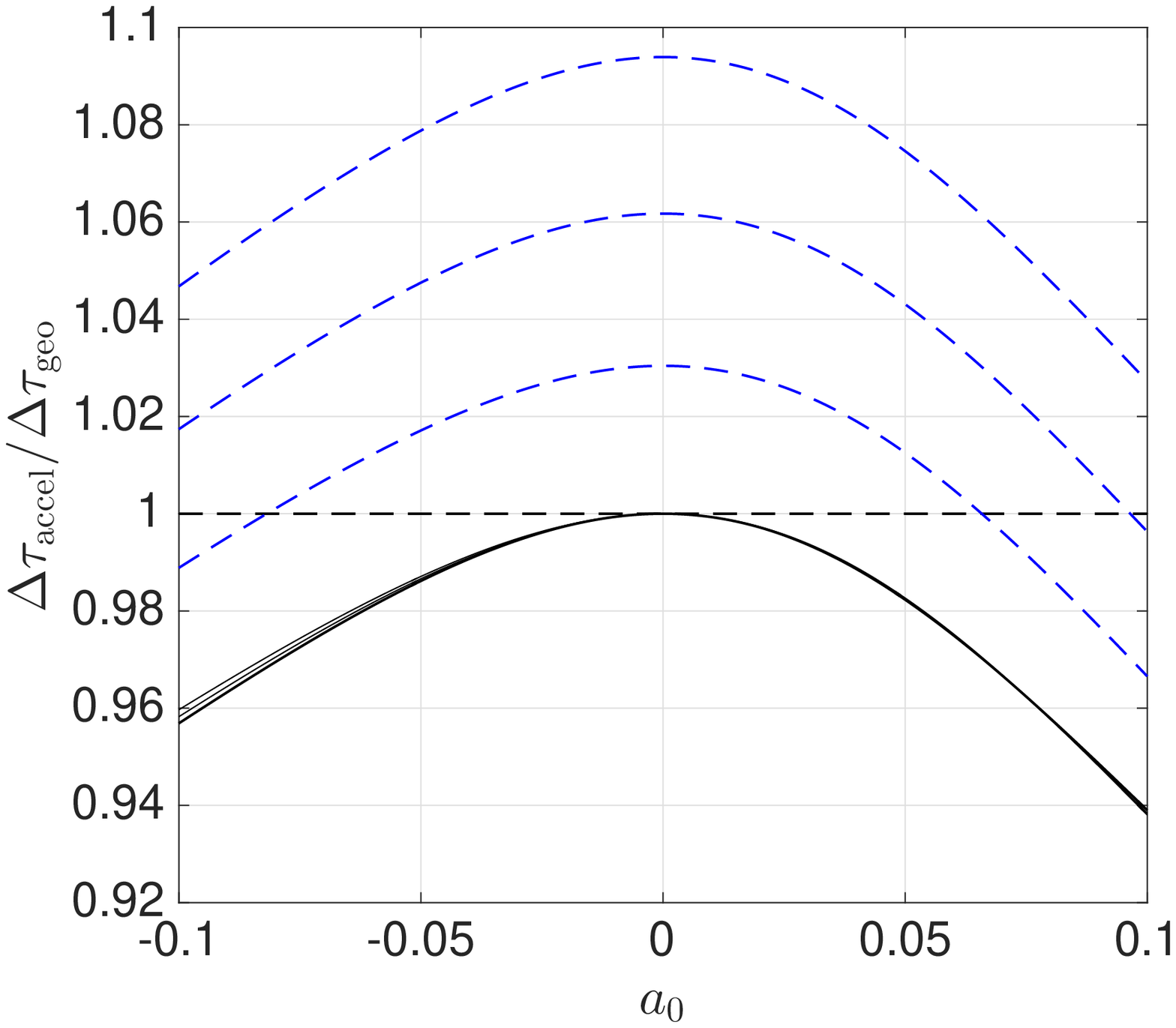}
\label{Schwarzaccel}
}
\caption{(a) A geodesic path (solid) and a sample of accelerated paths (dashed) within the Schwarzschild metric. The elapsed coordinate time was calculated using $M=1$, $r_0=6$, and $\sigma=0.55$. Time-like paths start at $r=6$, $\phi=\pi/2$, $\theta=\pi/2$, and end at $r=4$, $\phi=-\pi/4$, $\theta=\pi/2$. A range of values for $a_0$ were used: $a_0=\pm0.01,\pm0.05,\pm0.1$. (b) Proper time ratios for accelerated paths to geodesic paths within the Schwarzschild metric. The elapsed coordinate time was calculated using $M=1$, $r_0=6$, and $\sigma=0.55$. Time-like paths begin at $r=6$, $\phi=\pi/2$, $\theta=\pi/2$, and end at $r=4$, $\theta=\pi/2$. A range of final angular positions are explored: $\phi = 39\pi/30,41\pi/30,43\pi/30$. The solid lines correspond to azimuthally minimal geodesics, while the dashed lines correspond to azimuthally maximal geodesics.}
\end{figure}

By comparing the proper times of the two types of geodesic paths through spacetime, shown in Figure~\ref{Schwarzgeo}, we find again that the azimuthally minimal geodesic is longer for all paths investigated. Likewise, comparison of these two types of geodesics to a range of accelerated paths, as in Figure~\ref{BHaccelorbit} and \ref{Schwarzaccel}, provides the same ranking of experienced proper times as for the MT wormhole, in decreasing order as: azimuthally minimal geodesics, `weakly' accelerated paths, azimuthally maximal geodesics, and `strongly' accelerated paths.

%---------------------------------------------------------------------------------------------------------------

\section{Conclusion}

Examining the twin paradox in special relativity equips students with the idea that time is relative. However, it does not provide a complete description of the behaviour of time in the Universe. Analytically extending the problem to general relativity is often inaccessible to students due to its mathematical complexity. A computational approach provides simplicity, and is a nice tool to show students how objects behave in the metric under examination. The difficulty of the project may be tuned by simply choosing a metric of different complexity, allowing for the teaching of general relativity at an advanced level.

Here, we have described a computational method to numerically compute the age of a twin that passes between two given spacetime events. By application of this method to two simple metrics, we have shown that not only is the issue of proper time significantly more complex in general relativity, but also that seemingly contrary results may be found.  In opposition to the original paradox, accelerating twins may actually be the \textit{eldest} -- dependent upon the path taken by the other. However, while the absolutely longest path remains a geodesic, identifying that path from the possible geodesic paths remains complex; we leave this for a future contribution \cite{soko2}.

This method may be applied to any required metric, and is simple in its application. Consequently, it provides a unique means for students to investigate the behaviour of time under the effects of curvature, as well as to study individual metrics. For example, using this computational approach, we were able to reveal an apparent correlation between the traversed azimuthal angular distance of a geodesic and the experienced proper time (a result which we originally termed as the \emph{azimuthal hypothesis}). Although such a statement would need to be proven analytically, the quick formulation of the hypothesis demonstrates the utility of using this computational approach to explore and investigate the nature of time in general relativity. It may yet be seen that the azimuthally minimal path may be the globally longest geodesic in any spherically symmetric, static, stationary spacetime, but any such claim would require an exploration of further metrics than those undertaken here.

%---------------------------------------------------------------------------------------------------------------

\begin{ack}
The authors KKHF and HAC would like to acknowledge the Australian Postgraduate Awards (APA), through which this work was financially supported. HAC would additionally like to thank George F. R. Ellis for helpful correspondence. GFL gratefully acknowledges the Australian Research Council (ARC) for support through DP130100117.
\\
\end{ack}

%---------------------------------------------------------------------------------------------------------------

\newpage

%---------------------------------------------------------------------------------------------------------------

\appendix
\section{MATLAB Code}\label{appA}
\mbox{}\\
\textbf{Main file}\\
\begin{spacing}{0.9}
{\footnotesize
\texttt{
\\
\% This code solves the BVP to calculate the orbit and proper time of a particular twin.\\
{}\\
global M T b r0 rf phif a0 metric \% Declare global variables\\
{}\\
metric = 1; \% 1 for wormhole, 2 for blackhole\\
{}\\
M = 1; \% Mass of the black hole (M0)\\
b = 1; \% Wormhole throat parameter\\
r0 = 6; \% Initial radius\\
rf = 4; \% Final radius\\
phif = pi/4; \% Final angle\\
sigma = 0.6; \% Reunion time factor\\
a0 = 0.01; \% Acceleration parameter\\
{}\\
P = 2*pi*sqrt(r0$^\wedge$3/M); \% Period of a circular orbit\\
T = sigma*P; \% Elapsed coordinate time\\
t\_twin = linspace(0,T,250)'; \% Points to evaluate solution\\
{}\\
\% Guess initial values [tau r theta phi tau' r' theta' phi']\\
y\_init\_guess = [0 r0 pi/2 0 fn\_init(0,0) 0 0 0];\\
{}\\
\% Calculate solution\\
y\_init = bvpinit(t\_twin,y\_init\_guess); \% Initialise solution using initial guess\\
y\_solution = bvp4c(@(t,y) fn\_ode(t,y),@(ya,yb) fn\_bc(ya,yb),y\_init\_guess); \% BVP solver\\
y\_twin = (deval(y\_solution,t\_twin))'; \% Evaluate solution at chosen points\\
{}\\
\% Extract proper time\\
tau = t\_twin(end,1);\\
{}\\
\% Transform spherical to cartesian coordinates\\
x = y\_twin(:,2).*sin(y\_twin(:,3)).*cos(y\_twin(:,4)); \% x = rsin(theta)cos(phi)\\
y = y\_twin(:,2).*sin(y\_twin(:,3)).*sin(y\_twin(:,4)); \% y = rsin(theta)sin(phi)\\
{}\\
\% Plot results\\
plot(x,y);
}}
\end{spacing}

\newpage
\mbox{}\\
\textbf{Initial values}\\
\begin{spacing}{0.9}
{\footnotesize
\texttt{
\\
\% This code calculates the initial the component of the four-velocity, and assumes dtheta=0.\\
{}\\
function dtau = fn\_init(dr,dphi)\\
{}\\
global M b r0 metric\\
{}\\
if metric == 1; \% WH\\
\mbox{}\qquad dtau = sqrt(1 - dr$^\wedge$2 - (b$^\wedge$2 +r0$^\wedge$2)*dphi$^\wedge$2);\\
end\\
{}\\
if metric == 2; \% BH\\
\mbox{}\qquad dtau = sqrt((1-2*M/r0)-2*dr-r0$^\wedge$2*dphi$^\wedge$2);\\
end\\
{}\\
end\\
}}
\end{spacing}

\mbox{}\\
\textbf{Boundary Conditions}\\
\begin{spacing}{0.9}
{\footnotesize
\texttt{
\\
\% This code sets the boundary conditions for the boundary value problem.\\
\% ya(i) is to the boundary condition of the i-th variable at t = 0.\\
\% yb(i) is the boundary condition of the i-th variable at t = T.\\
{}\\
function res = fn\_bc(ya,yb)\\
{}\\
global r0 rf phif\\
{}\\
res = [ya(1) ya(2)-r0 ya(3)-pi/2 ya(4)-pi/2 ya(5)-fn\_init(ya(6),ya(8)) yb(2)-rf yb(3)-pi/2 yb(4)-phif];\\
{}\\
end\\
}}
\end{spacing}
\newpage

\mbox{}\\
\textbf{Equations of Motion}\\
\begin{spacing}{0.9}
{\footnotesize
\texttt{
\\
\% This code contains the equations of motion that are integrated.\\
{}\\
function dy = fn\_ode(t,y)\\
{}\\
global M T b a0 metric\\
{}\\
dy = zeros(8,1);\\
{}\\
ar = [4*a0/T*abs(t-T/2)-a0,0,0]; \% Four-acceleration spatial components\\
{}\\
if metric == 1  \% Morris Thorne wormhole metric\\
\mbox{}\qquad \% Define Christoffel symbols\\
\mbox{}\qquad \% Time component\\
\mbox{}\qquad C\_t = zeros(4,4);\\
\mbox{}\qquad \% Spatial components\\
\mbox{}\qquad C\_i = zeros(4,4,3);\\
\mbox{}\qquad C\_i(3,3,1) = -y(2);\\
\mbox{}\qquad C\_i(4,4,1) = -y(2)*sin(y(3))$^\wedge$2;\\
\mbox{}\qquad C\_i(2,3,2) = y(2)/(b$^\wedge$2+y(2)$^\wedge$2);\\
\mbox{}\qquad C\_i(3,2,2) = y(2)/(b$^\wedge$2+y(2)$^\wedge$2);\\
\mbox{}\qquad C\_i(4,4,2) = -sin(y(3))*cos(y(3));\\
\mbox{}\qquad C\_i(2,4,3) = y(2)/(b$^\wedge$2+y(2)$^\wedge$2);\\
\mbox{}\qquad C\_i(4,2,3) = y(2)/(b$^\wedge$2+y(2)$^\wedge$2);\\
\mbox{}\qquad C\_i(3,4,3) = cot(y(3));\\
\mbox{}\qquad C\_i(4,3,3) = cot(y(3));\\
\mbox{}\qquad \% Define metric\\
\mbox{}\qquad gtt = -1;\\
\mbox{}\qquad grr = 1;\\
\mbox{}\qquad gthetatheta = b$^\wedge$2 + y(2)$^\wedge$2;\\
\mbox{}\qquad gphiphi = (b$^\wedge$2 + y(2)$^\wedge$2)*sin(y(3))$^\wedge$2;\\
\mbox{}\qquad \% Define a$^\wedge$t for given metric (simplified, as metric is diagonal)\\
\mbox{}\qquad at = -(grr*y(6)*ar(1) + gthetatheta*y(7)*ar(2) + gphiphi*y(8)*ar(3))/gtt;\\
end\\
if metric == 1  \% Schwarzschild metric\\
\mbox{}\qquad \% Define Christoffel symbols:\\
\mbox{}\qquad \% Time component\\
\mbox{}\qquad C\_t = zeros(4);\\
\mbox{}\qquad C\_t(1,1) = M/y(2)$^\wedge$2;\\
\mbox{}\qquad C\_t(3,3) = -y(2);\\
\mbox{}\qquad C\_t(4,4) = -y(2)*sin(y(3))$^\wedge$2;\\
\mbox{}\qquad \% Spatial components\\
\mbox{}\qquad C\_i = zeros(4,4,3);\\
\mbox{}\qquad C\_i(1,1,1) = M*(y(2) - 2*M)/y(2)$^\wedge$3;\\
\mbox{}\qquad C\_i(1,2,1) = -M/y(2)$^\wedge$2;\\
\mbox{}\qquad C\_i(2,1,1) = -M/y(2)$^\wedge$2;\\
\mbox{}\qquad C\_i(3,3,1) = -y(2)+2*M;\\
\mbox{}\qquad C\_i(4,4,1) = -(y(2)-2*M)*sin(y(3))$^\wedge$2;\\
\mbox{}\qquad C\_i(3,2,2) = 1/y(2);\\
\mbox{}\qquad C\_i(2,3,2) = 1/y(2);\\
\mbox{}\qquad C\_i(4,4,2) = -sin(y(3))*cos(y(3));\\
\mbox{}\qquad C\_i(4,2,3) = 1/y(2);\\
\mbox{}\qquad C\_i(2,4,3) = 1/y(2);\\
\mbox{}\qquad C\_i(4,3,3) = cot(y(3));\\
\mbox{}\qquad C\_i(3,4,3) = cot(y(3));\\
\mbox{}\qquad \% Define metric\\
\mbox{}\qquad gtt = -(1-2*M/y(2));\\
\mbox{}\qquad gthetatheta = y(2)$^\wedge$2;\\
\mbox{}\qquad gphiphi = y(2)$^\wedge$2*sin(y(3))$^\wedge$2;\\
\mbox{}\qquad \% Define a$^\wedge$t for given metric\\
\mbox{}\qquad at = -(ar(1) + gthetatheta*y(7)*ar(2) + gphiphi*y(8)*ar(3))/(gtt + y(6));\\
end\\
{}\\
\% Redefine velocity vector so that y(1) = dtau/dtau = 1\\
y2 = y;\\
y2(5) = 1;\\
\% Set time components of equation of motion\\
dy(1) = y(5);\\
dy(5) = y(5)*y2(5:8)'*C\_t*y2(5:8) - at*y(5)$^\wedge$3;\\
\% Set spatial components of equations of motion\\
for i = 1:3\\
\mbox{}\qquad dy(i+1) = y(i+5);\\
\mbox{}\qquad dy(i+5) = y(i+5)*y2(5:8)'*C\_t*y2(5:8) - y2(5:8)'*C\_i(:,:,i)*y2(5:8) + y(5)$^\wedge$2*(ar(i) \qquad - at*y(i+5));\\
end\\
%\mbox{}\qquad dy(i+1) = y(i+5);\\
%\mbox{}\qquad dy(i+5) = y(i+5)*y2(5:8)'*C_t*y2(5:8) - y2(5:8)'*C_i(:,:,i)*y2(5:8) + y(5)^2*(ar(i) - at*y(i+5));\\
%end\\
}}
\end{spacing}

\section*{References}
\begin{thebibliography}{16}
\bibitem{frisch} Frisch O 1962 Time and relativity: Part II {\it Contemp. Phys.} \textbf{3} 194--201
\bibitem{builder} Builder G 1959 Resolution of the clock paradox {\it Am. J. Phys.} \textbf{27} 656--658
\bibitem{scott} Scott G D 1959 On solutions of the clock paradox {\it Am. J. Phys.} \textbf{27} 580--584
\bibitem{swift} Holstein B R and Swift A R 1972 The relativity twins in free fall {\it Am. J. Phys.} \textbf{40} 746--750
\bibitem{abram1} Abramowicz M A and Bajtlik S 2009 Adding to the paradox: The accelerated twin is older arXiv:0905.2428 [physics.class-ph]
\bibitem{abram2} Abramowicz M A, Bajtlik S and Klu\'{z}niak W 2007 The twin paradox on the photon sphere {\it Phys. Rev. A} \textbf{75} 044101
\bibitem{gron} Gr\o n~\O~and Braeck S 2011 The twin paradox in a cosmological context {\it Eur. Phys. J. Plus} \textbf{126} 79--92
\bibitem{soko} Soko{\l}owski L 2012 On the twin paradox in static spacetimes: I. Schwarzschild metric {\it Gen. Relativ. Gravit.} \textbf{44} 1267--1283
\bibitem{markley} Markley F L 1973 Relativity twins in the Kerr metric {\it Am. J. Phys.} \textbf{41} 1246--1250
\bibitem{mt} Morris M and Thorne K 1988 Wormholes in spacetime and their use for interstellar travel: A tool for teaching general relativity {\it Am. J. Phys.} \textbf{56} 395--412
\bibitem{catalogue} Mueller T and Grave F 2009 Catalogue of spacetimes arXiv:0904.4184 [gr-qc].
\bibitem{soko2} Soko{\l}owski L M and Golda Z A 2015 Jacobi fields and conjugate points on timelike geodesics in special spacetimes {\it Acta Physica Polonica B} \textbf{46} 773--787
\end{thebibliography}
\end{document}